\documentclass[twocolumn,pra,amsmath,amssymb,superscriptaddress]{revtex4-1}
\usepackage[utf8]{inputenc}

\usepackage{amsfonts,amsmath,amssymb}
\usepackage{siunitx}
\usepackage{graphicx}
\usepackage{hyperref}
\usepackage{bm} % boldmath
\usepackage{mathrsfs}
\usepackage{textcomp}
\usepackage{subfigure}
\newcommand{\degree}{\ensuremath{^\circ}}%

\newcommand{\mean}[1]{{ \langle{#1} \rangle}}

\newlength{\figwidth}
\newlength{\figwidthsmall}
\begin{document}

\title{Control and femtosecond time-resolved imaging of torsion in a chiral molecule}

\author{Jonas L. Hansen}%
\affiliation{Interdisciplinary Nanoscience Center (iNANO), Aarhus University, DK-8000 Aarhus C,
  Denmark}
 \author{Jens H. Nielsen}%
\affiliation{Department of Physics and Astronomy, Aarhus University, DK-8000 Aarhus C, Denmark}
\author{Christian Bruun Madsen}
\affiliation{J. R. Macdonald Laboratory, Kansas State University, Manhattan, Kansas, 66506, USA}
\author{Anders Thyboe Lindhardt}
\affiliation{Interdisciplinary Nanoscience Center (iNANO), Aarhus University, DK-8000 Aarhus C,
  Denmark}
\affiliation{Department of Chemistry, Aarhus University, DK-8000 Aarhus C, Denmark}
\author{Mikael P. Johansson}
\affiliation{ Institut de Química Computacional and Departament de Química, Universitat de Girona, Campus Montilivi, ES-17071, Girona, Spain}
\author{Troels Skrydstrup}
\affiliation{Interdisciplinary Nanoscience Center (iNANO), Aarhus University, DK-8000 Aarhus C,
  Denmark}
\affiliation{Department of Chemistry, Aarhus University, DK-8000 Aarhus C, Denmark}
\author{Lars Bojer Madsen}
\affiliation{Lundbeck Foundation Theoretical Center for Quantum System Research, Department of Physics and Astronomy, Aarhus University, DK-8000 Aarhus C, Denmark}
\author{Henrik Stapelfeldt}%
\affiliation{Interdisciplinary Nanoscience Center (iNANO), Aarhus University, DK-8000 Aarhus C,
  Denmark}
\affiliation{Department of Chemistry, Aarhus University, DK-8000 Aarhus C, Denmark}
\email[Corresponding author: ]{henriks@chem.au.dk}%

\date{\today}

\begin{abstract}
 % In this paper we provide new experimental insights on how the combination of long (nanosecond) and short (femtosecond) laser pulses can be used to induce torsional motion in the axially chiral biphenyl derivative 3,5-difluoro-3´,5´-dibromo-4´-cyanobiphenyl. The long elliptically polarized laser pulse serves to spatially align the most and second most polarizable axis of the molecule, whereas the linearly polarized short pulse initiates torsion about the stereogenic axis. Experimentally the induced torsional motion can be monitored directly by extracting the dihedral angle using femtosecond time-resolved Coulomb explosion. At short kick-probe time delays (0--4~ps) the experimentally observed torsional motion of the dihedral angle (amplitude of 3 degrees) can be modeled by theoretical calculations which assume that the motion of the two phenyl rings are independent \cite{Madsen:PRL:2009}.  At longer time delays ($\geq$4~ps) the two phenyl-planes, are observed to have different dynamics and delocalize at different times, which is also fully compatible with the existing theory.
 % Furthermore, a new imaging analysis technique relying on the correlation between the ejected ionic fragments supports our interpretation of the experimental data.

We study how the combination of long and
  short %, on the molecular rotational time-scale,
  laser pulses, can be used to induce torsion in an axially chiral biphenyl derivative (3,5-difluoro-3´,5´-dibromo-4´-cyanobiphenyl). A long, with respect to the molecular rotational periods,
  elliptically polarized laser pulse produces 3D alignment of the molecules, and a linearly polarized short pulse initiates torsion about the
  stereogenic axis. The torsional motion is monitored in real-time by
  measuring the dihedral angle using femtosecond time-resolved Coulomb explosion imaging.
  %The experimental setup allows us to measure the dihedral angle with unprecedented precision and thereby trace the torsional motion up to several picoseconds.
  %The measurements reveal new details on the torsional dynamics and its coupling to the overall rotation of the molecule.
  Within the first 4 picoseconds, torsion occurs with a period of 1.25 picoseconds and an amplitude of 3$^\circ$ in excellent agreement with theoretical calculations.
  At larger times the quantum states of the molecules describing the torsional motion dephase and an almost isotropic distribution of the dihedral angle is measured.
  %At longer times torsion is blurred by delocalization of the molecular orientation due to overall rotation of the molecule consistent with our theoretical model.
  We demonstrate an original application of covariance analysis of two-dimensional ion images to reveal strong correlations between specific ejected ionic fragments from Coulomb explosion. This technique strengthens our interpretation of the experimental data.
  % The theoretical analysis accounts for and generalizes the experimental findings ... and on the
  % basis of these results we discuss future applications of laser-induced torsion, time-resolved
  % studies of deracemization and laser controlled molecular junctions based on molecules with
  % torsion.
\end{abstract}

\pacs{}

\maketitle

\section{Introduction}
The control of molecules and chemical reactions with lasers is one of the main goals of femtochemistry \cite{zewail:science:1988}. One particular aspect of such studies is
the use of laser pulses to control the transition from one enantiomer of a chiral molecule to its mirror form. This topic has been the subject of a large number of theoretical studies and these efforts are motivated by the intriguing prospect of light-induced deracemization \cite{Shoa:JCP:1997,Fujimura:cpl:1999,Gerbasi:JCP:2001,Kral:PRL:2001,hoki_control_2002,Kroner:cp:2003,Kroner:CP:2007}, i.e., creation of enantiomeric excess.
Axially chiral molecules have attracted special interest because the potential barrier separating an enantiomeric pair is often low so that lasers at moderate intensities may be applied to facilitate efficient crossing of the barrier without fragmenting the molecule. For axially chiral molecules the stereogenic element is an axis (connecting two atoms) rather than a point (a single atom) and the reaction path separating the two enantiomers corresponds to torsion around the stereogenic axis \cite{Eliel_Wilen_1994}.

The first theoretical light-based scheme which successfully showed the production of enanotiomeric excess ($50.0001\%:49.9999\%$) involved circularly polarized continuous wave monochromatic light \cite{Shoa:JCP:1997}.
However, a much higher conversion rate is needed if the deracemization is to be confirmed experimentally - not to mention to have practical applications. To this end studies showed that almost perfect deracemization could be achieved by exposing
phosphinothioic acid (H$_2$OPSH) to picosecond terahertz laser pulses \cite{Fujimura:cpl:1999}.
In these studies, the key was to realize that changes in the torsional coordinate was associated with a shift in the permanent dipole moment, a feature that was exploited by using polarization shaped laser pulses to transfer a racemate into the desired enantiomeric form.
%where the difference in field--dipole interaction for the two enantiomers could be exploited by choosing the optimal electrical field polarization to achieve almost perfect deracemization. In particular if changes in the torsional coordinate is associated with a shift in the direction of the permanent dipole moment, then a racemate can be transferred into the desired enantiomeric form by controlling the polarization of the laser pulse \cite{Fujimura:cpl:1999}.
Since this discovery, several different approaches have been proposed to control torsional motion in axially chiral molecules. These approaches include the use of coherent control or phase controlled pulses \cite{Shapiro:JCP:1991} to perform, e.g., \textit{laser distillation} \cite{Gerbasi:JCP:2001} where excitation of one of the enantiomers can be enhanced
%with phase controlled pulses through symmetry-breaking interaction
by coupling the ground state to a superposition of two excited states with opposite parities, or by \textit{cyclic population transfer} that relies on quantum interference in an effective three-level-system \cite{Kral:PRL:2001}. Alternatively, the deracemization can be achieved by exploiting that the direction of the electronic transition moment vector to an excited state is different for the two enantiomers. One way to exploit this directionality is a \textit{pump-dump} scheme \cite{hoki_control_2002}, where the population is gradually transferred from one enantiomer to the other. Another method would be \textit{asymmetric excitation}, where the excited state is repulsive and leads to dissociation thereby removing this enantiomeric form from the ensemble \cite{Kroner:cp:2003}.
A common feature for most of these schemes is that they either require or benefit from pre-aligned and oriented molecular targets \cite{Kroner:CP:2007}.

Since alignment and orientation is generally required in the enantiomer-control-schemes a more straightforward, and thereby experimentally feasible approach, would be to rely on alignment concepts alone to control the torsion.
Recently, this has been the focus of several studies.
These studies fall into two regimes: (i) the adiabatic regime where the pump laser that induces torsion is long compared to the time scale of torsional dynamics \cite{Seideman:prl:2007,Coudert:PRL:2011,Parker:JCP:2011} and (ii) torsion induced by a short non-adiabatic pump pulse \cite{Madsen:PRL:2009, Madsen:JCP:2009}.
For those axially chiral molecules where the stereogenic axis and the most polarizable axis coincide it was suggested that torsion could be controlled adiabatically by a single elliptically polarized nanosecond (ns) laser pulse provided that the torsional barrier is sufficiently low. This pulse potentially serves to both induce 3D alignment and torsional motion by interaction of the minor axis of the polarization ellipse with the polarizability components of the two moieties forcing them into coplanarity \cite{Seideman:prl:2007,Parker:JCP:2011}. The feasibility of this approach has recently been disputed due to the coupling between overall rotation and torsional motion leading to a breakdown of torsional alignment \cite{Coudert:PRL:2011}.

So far only very few experimental papers have investigated the possibility to control and do real-time monitoring of the torsion in axially chiral molecules.
In Refs.~\cite{Madsen:PRL:2009, Madsen:JCP:2009} the stereogenic axis of a 3,5-difluoro-3´,5´-dibromo-biphenyl molecule was held fixed-in-space  by 1-dimensional (1D) adiabatic alignment with a linearly polarized 9 ns long laser pulse. In equilibrium the angle between the F-phenyl plane and the Br-phenyl plane is $\pm39\degree$. Upon irradiation with a 700 femtosecond (fs) long, intense (but non-fragmenting), nonresonant pump pulse, torsion was induced accompanied by overall rotation around the fixed stereogenic axis. The torsional and rotational dynamics, measured by inducing Coulomb explosion with a short intense delayed probe pulse and recording of the emission direction of recoiling Br$^+$ and F$^+$ ions through ion imaging, showed that the amplitude and period of torsion was 0.6$\degree$ and 1 picosecond (ps).
While the period was in qualitative agreement with the theoretical model prediction of 1.2~ps, the amplitude predicted by theory was larger, 2.45$\degree$. The theory rationalized that stimulated Raman transitions, driven by the pump pulse, creates a wave packet of torsional eigenstates in the electronic ground state leading to the torsion observed. It is one of the aims of the present work to resolve the discrepancy between theory and experiment.

The principle of the current work is similar to the previous work, \cite{Madsen:PRL:2009, Madsen:JCP:2009} but three key experimental factors were changed to achieve much better resolved peaks in the experimental ion images which allows for tracing of the torsional motion with unprecedented precision and, therefore, a quantitative rather than a qualitative comparison to the calculated results.
The three decisive factors are: 1) A chemically much purer sample resulting in higher contrast ion images. 2) The molecules were 3D aligned \cite{larsen:2000:prl,Nevo:PCCP:2009} prior to the kick and the probe pulses, i.e., the entire molecule was fixed-in-space not just the stereogenic axis (1D alignment). This implies that the torsional motion stands out much more distinctly in the ion images. 3) The molecule employed here, 3,5-difluoro-3´,5´-dibromo-4´-cyanobiphenyl, is almost identical to the molecule studied in Refs.~\cite{Madsen:PRL:2009, Madsen:JCP:2009} except that it has a nitrile group attached to the end of the Br-phenyl. This gives the current molecule a dipole moment of 4.4 Debye
    \cite{Note1}
    %\footnote{Dipole moments and polarizabilities were computed with {\sc Turbomole} 6.2 \cite{Turbomole} at density functional theory (DFT) level, employing the B3LYP functional which combines Becke's three-parameter hybrid exchange functional \cite{B3} with the Lee-Yang-Parr correlation functional \cite{LYP}; the correlation of the uniform electron gas was modeled with the Vosko-Wilk-Nusair VWN5 formulation \cite{VWN}. The doubly-polarized triple-zeta basis set, TZVPP, \cite{TZVPP} was used throughout. This level of theory has been shown to be suitable for the study of biphenyls \cite{J&O}}
compared to 0.2 Debye \cite{Note1} for the molecule used previously.
The large permanent dipole moment of the molecules made it possible to use an electrostatic deflector to spatially separate molecules according to their rotational quantum state \cite{Holmegaard:PRL102:023001,Filsinger:JCP:2009}. Performing the separation prior to the interaction with the laser pulses, allowed us to focus the laser pulses onto only the molecules with low rotational energy, improving the degree of prealignment \cite{Holmegaard:PRL102:023001,Filsinger:JCP:2009}.
 %to quantum state-select them, by an electrostatic deflector \cite{Holmegaard:PRL102:023001,Filsinger:JCP:2009}, prior to the interaction with the laser pulses. This serves to lower the average rotational energy and improve the degree of prealignment.

The paper is organized as follows. In Sec. \ref{Experimental Methods}, the experimental setup is described. In Sec. \ref{Experimental Results} the results are discussed and Sec. \ref{theory} provides a comparison with theory. Section \ref{conclusions} concludes.

\section{Experimental Setup}
\label{Experimental Methods}

The experimental setup has recently been described in detail \cite{Hansen:PRA:2011} and
only a brief description of the most important features is given here.
A cold molecular beam ($\sim$1K) is formed by heating less than fifty milligram of solid 3,5-difluoro-3´,5´-dibromo-4´-cyanobiphenyl in 90 bar of He to 170$^\circ$C and expanding the mixture into vacuum through a pulsed Even-Lavie valve. This chemical is not commercially available and was synthesized specially for the experiment (see Appendix).
Here it should be noted that in the previously reported experiment the molecular sample was slightly contaminated which can be seen directly from close inspection of the NMR spectrum in Fig. 4 (d) of \cite{Madsen:JCP:2009}. In particular the NMR spectrum shows a small impurity (5\%) of the biproduct where both phenyl rings have flourine substituents. This contamination is problematic since the vapor pressure of the biproduct is higher than the target molecule and, therefore, the ratio between impurity and target molecules will be higher in the molecular beam than the initial 5\% in the powder sample. Furthermore, the second most polarizable axis of the impurity will be located halfway between the two phenyl rings and, therefore, contribute to significant blurring of the F$^+$ ion images and, thus, reduce the capability to experimentally resolve the torsional motion.
%
%A cold molecular beam ($\sim$1K) is formed by expanding X~mbar 3,5-difluoro-3´,5´-dibromo-4´-cyanobiphenyl seeded in 90 bar of He into vacuum through a pulsed Even-Lavie valve heated to 170$^\circ$C.
%

\begin{figure}
    \centering
   \includegraphics[width=0.45\textwidth]{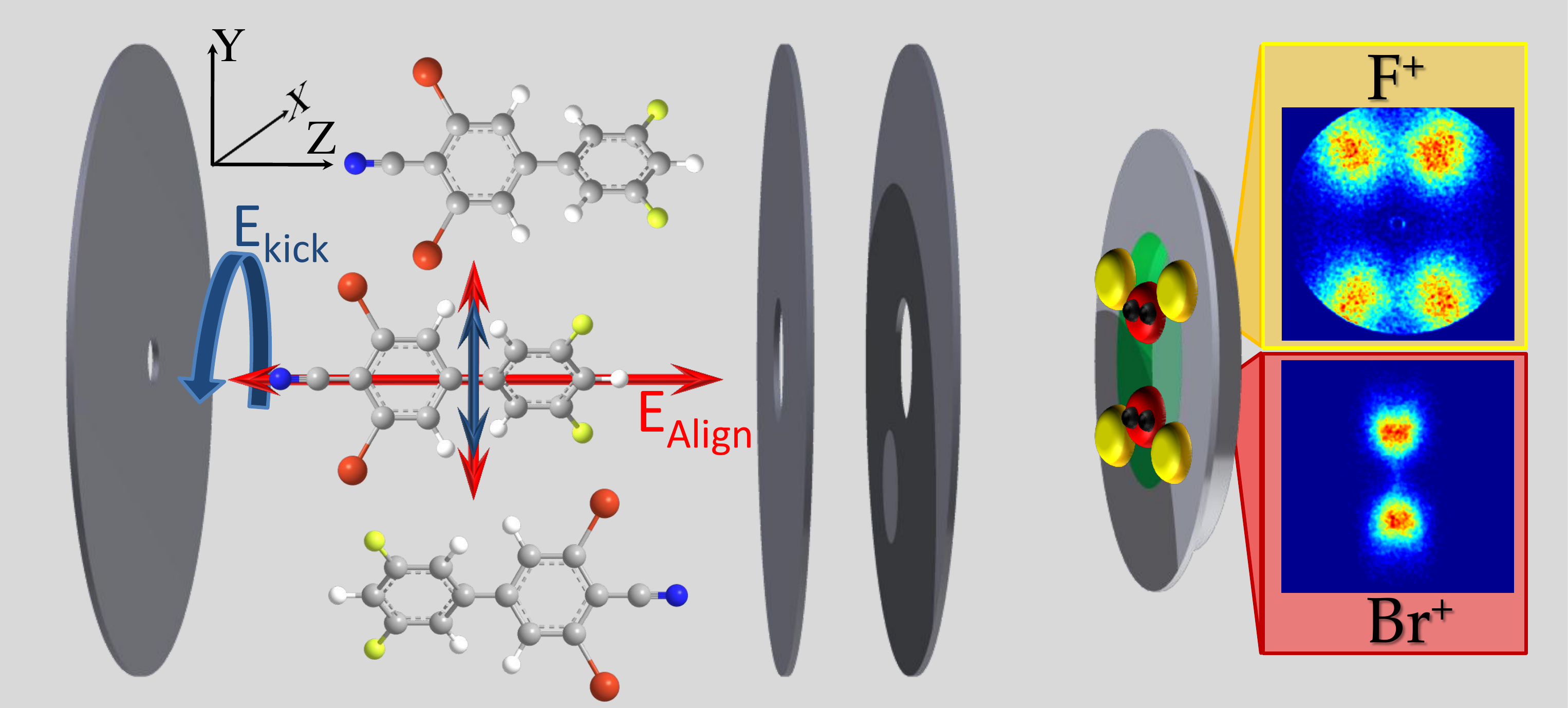}
   \caption{ Schematic of the velocity map imaging spectrometer used to detect F$^+$ and Br$^+$
     ions. For illustrative purposes, the regions on the detector screen where F$^+$ (Br$^+$) ions are expected to hit for perfectly aligned molecules, at equilibrium, are marked in yellow (black). The images behind the detector are experimental data recorded 2 ps after the kick-pulse. The polarization state of the YAG (alignment) pulse ($\lambda = 1064~\textup{nm}$, $\tau_\text{FWHM} = 10$~ns, I$_{\textup{YAG}}$ = \SI{6e11}{\watt/\centi\meter\squared}) and the kick pulse (800~nm, 200~fs, I$_{\textup{kick}}$ = \SI{2e13}{\watt/\centi\meter\squared}) are also displayed.
    % The static electric field of the spectrometer, pointing from the repeller to the extractor electrode for ion detection, breaks the head-for-tail symmetry in a quantum state selected sample by     preferentially placing the CN-end towards the repeller where the positive voltage is highest.
    }
  \label{fig:Schematic_setup}
\end{figure}

The beam of cold molecules pass through an electrostatic deflector that deflects the molecules according to their effective dipole moment. The effective dipole moment is determined by its rotational quantum state \cite{Filsinger:JCP:2009}.
%determined by the rotational quantum state occupied by the individual molecules \cite{Filsinger:JCP:2009}.
Following deflection, the molecules enter a velocity map imaging (VMI) spectrometer (\autoref{fig:Schematic_setup}) where they are
irradiated by three pulsed laser beams. The first pulse (YAG pulse: $\lambda = 1064~\textup{nm}$,
$\tau_\text{FWHM} = 10$~ns, I$_{\textup{YAG}}$ = \SI{6e11}{\watt/\centi\meter\squared}) serves to adiabatically align the molecules. It is elliptically polarized with the major axis parallel to the spectrometer axis, the Z-axis (see \autoref{fig:Schematic_setup}). The second pulse (kick
pulse: 800~nm, 200~fs, I$_{\textup{kick}}$ = \SI{2e13}{\watt/\centi\meter\squared}) initiates torsion and it is linearly polarized  along the Y-axis. The third pulse (probe pulse: 800~nm, 30~fs, I$_{\textup{probe}}$ = \SI{3e14}{\watt/\centi\meter\squared}) is used to characterize the spatial
orientation of the two phenyl planes by Coulomb exploding the molecules and recording the recoiling
F$^+$ and Br$^+$ fragments. It is linearly polarized either along the Z-axis or the Y-axis. By recording series of images for different kick-probe delays, $t$,
the evolution of the torsional motion can be monitored.  To ensure that the molecules probed are both aligned and torsionally excited, the spotsize of the three
laser pulses are adjusted such that the foci of the YAG and kick pulses are large ($\omega_{0} =
\SI{38}{\um}$ and $\omega_{0} = \SI{35}{\um}$) compared to the probe pulse
($\omega_{0} =\SI{26}{\um}$).

%It should also be noted that the spatial dispersion of the molecular beam by the deflector allows for quantum state selection in the VMI spectrometer simply by moving the foci of the laser pulses throughout the molecular beam. {\bf Maybe add sentence that we expect the molecules in the deflected region to align and orient better.}

%YAG: $\omega_{0} = 37.6\mu m ; 10 \textrm{ns} ; I_{YAG} \sim 6 \times 10^{11} W cm^{-2}$
%Kick: $\omega_{0} = 35.4\mu m ; 200 \textrm{fs} ; I_{kick} \sim 2 \times 10^{13} W cm^{-2}$
%Cou: $\omega_{0} = 26.3\mu m ; 30 \textrm{fs} ; I_{Cou} \sim 3 \times 10^{14} W cm^{-2}$

%\begin{figure}
%   \centering
%   \includegraphics[width=0.8\textwidth]{1_Vacuum_chamber_v2}
%   \caption{(Color online) Exploded view of the molecular beam machine consisting of (from the
%     right) the source chamber, the deflector chamber and the target chamber. In this experiment
%     three laser beams are used, the 1064 nm YAG beam (blue cylinder) providing adiabatic alignment
%     of the molecules, an 800 nm femtosecond beam (green cylinder) to induce rotational/torsional
%     dynamics and the 800 nm probe beam (red cylinder) used for coulomb explosion.  The pair of
%     circular discs in each laser beam represents a half-wave and a quarter-wave plate (closest to
%     the vacuum chamber) used to control the polarization state of the laser pulses.}
%   \label{fig:vacuum-chamber}
%\end{figure}

\section{Experimental Results}
\label{Experimental Results}

\subsection{Deflection}
\label{deflection}

\begin{figure}
   \centering
   \includegraphics[width=0.45\textwidth]{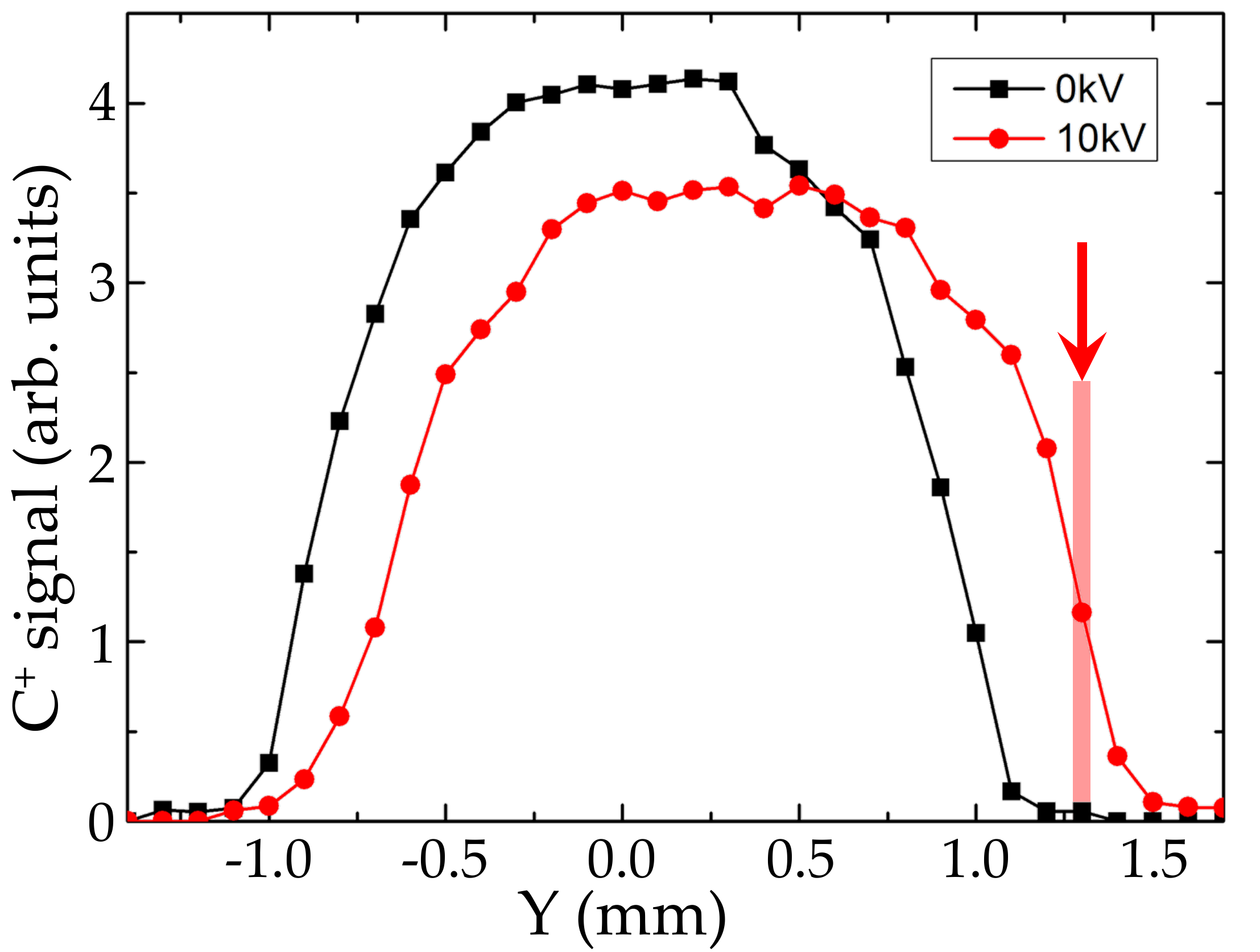}
   \caption{ Vertical profiles of the molecular beam measured by recording the C$^{+}$
     signal as a function of the vertical position of the probe beam focus. The experimental data
     are shown by black squares (deflector off, 0~kV) and red circles (deflector on, 10~kV). The red arrow at Y~=~1.3~mm indicates the position of the laser foci used to acquire ion images of
     % undeflected (black), Y~=~0~mm, and
     the deflected molecules.}
   \label{fig:deflection}
\end{figure}

The effect of the deflector on the molecular beam is shown in \autoref{fig:deflection} by the
vertical intensity profiles. They are obtained by recording the magnitude of an ion signal, produced by the probe pulse only, as a function of the vertical position of the laser focus similar to what was previously reported (see, for instance, Refs.~\cite{Holmegaard:PRL102:023001,Nevo:PCCP:2009,Hansen:PRA:2011}).
Here, we use the strongest peak in the ion time-of-flight-spectrum which is the C$^{+}$ signal.
When the deflector is off, the molecular beam extends over $\sim$1.5 mm, mainly determined by the diameter of
the skimmer before the deflector. When the deflector is turned on, the molecular beam profile
broadens and shifts upwards (positive Y-values). The molecules in the lowest rotational quantum states
have the largest effective dipole moments and are, therefore, deflected the most, as shown in recent work on iodobenzene and benzonitrile molecules~\cite{Holmegaard:PRL102:023001,Filsinger:JCP:2009}. In the measurements described below, experiments were conducted on quantum state-selected molecules simply by positioning the laser foci close to the upper cut-off region in the 10~kV profile, indicated by the red arrow in \autoref{fig:deflection} (Y~=~1.3~mm). Our motivation for using the deflected part of the beam is that despite the density of molecules is lower, the alignment is higher \cite{Holmegaard:PRL102:023001,Filsinger:JCP:2009,Nevo:PCCP:2009}.

\subsection{Adiabatic prealignment}
\label{alignment-orientation}
The 3,5-difluoro-3´,5´-dibromo-4´-cyanobiphenyl (C$_{13}$H$_{5}$F$_{2}$Br$_{2}$N, to
be denoted by DFDBrCNBph in the following) is a chiral asymmetric top molecule, and the molecular composition in the supersonic expansion is a racemate, i.e., consisting of 50\% of each of the two enantiomers.

Three-dimensional alignment of the molecules is achieved using an elliptically polarized YAG pulse
which confines both the most polarizable axis (MPA) and the second most polarizable axis
(SMPA) of the molecule %\footnote{The dynamic polarizabilities of the molecule, at equilibrium, $\lambda$ = 800 nm are $\alpha_{zz}= 46.83~\textrm{Å}^{3}$, $\alpha_{yy}= 16.82~\textrm{Å}^{3}$, $\alpha_{xx}= 32.02~\textrm{Å}^{3}$} along the major and minor axis of the polarization ellipse
\cite{Note2,stapelfeldt:2003:rmp,larsen:2000:prl,Nevo:PCCP:2009}, respectively. In DFDBrCNBph the MPA is
located along the stereogenic axis (C-C bond axis connecting the two phenyl rings -- see
\autoref{fig:Schematic_setup}) whereas the SMPA is perpendicular to this at an angle of
$10\degree$ from the phenyl-plane with the nitrile and bromine substituents towards the other
phenyl ring (see \autoref{fig:Bp_angle_evo_short}). The effect of applying the YAG pulse, having its
major and minor polarization axis along the Z-axis and the Y-axis, respectively, can be seen in panels (a) and (b) at $t = -0.33$ps in \autoref{fig:Bp_ion_evo_short}, showing F$^+$ and Br$^+$ ion images from Coulomb exploding
the molecules at the peak of the YAG pulse.  In both images two distinct features are
observed. First, the center of the detector is almost void of ions (a minor region has been cut in
the F$^+$ images to exclude a noise signal from residual water molecules in the chamber). This observation is only compatible
with the stereogenic axis (C-C bond axis) being confined perpendicular to the detector
plane, i.e., along the Z-axis.
Secondly, it is seen that both the Br$^+$ and the F$^+$ ions localize around the minor polarization axis of the YAG pulse (the Y-axis), the confinement being more pronounced for the Br$^+$ ions. We interpret this as the SMPA of the molecule being aligned along the Y-axis. The more spread-out signal of the F$^+$ ions results from the fact that the F-phenyl ring, at equilibrium, is offset 28\degree~ from the SMPA whereas the Br-phenyl ring is only offset by 10\degree. Thus, the joint observations of the F$^+$ and Br$^+$ images demonstrate that the molecules are 3D aligned as expected.

\subsection{Time-dependent Torsion}
\label{exp_data}

\begin{figure*}
    \centering
   \includegraphics[width=1.0\textwidth]{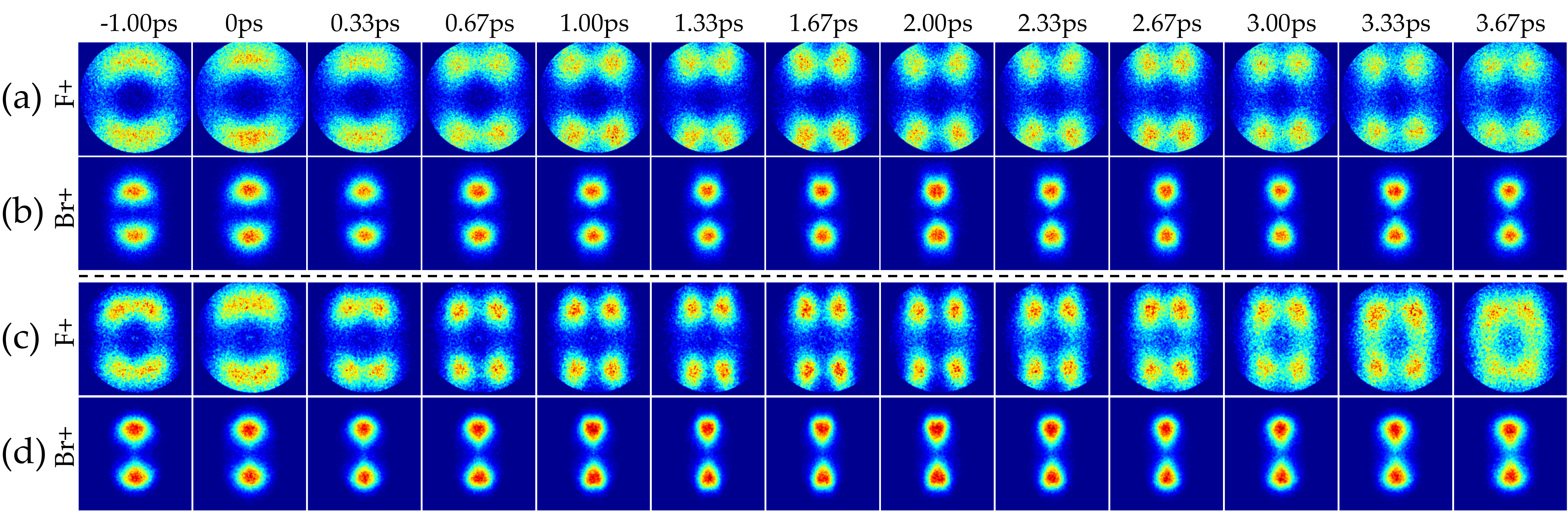}
   \caption{ F$^+$ (panel (a) and (c)) and Br$^+$ (panel (b) and (d)) ion images obtained as a function of the kick-probe time delay indicated in ps above each column of panels. Panels (a) and (b) ((c) and (d)) were obtained using a linearly polarized probe pulse being perpendicular to (parallel with) the detector.
   The delay of the probe pulse ($\lambda = 800~\textup{nm}$, $\tau_\text{FWHM} = 30~\textup{fs}$, I$_{\textup{probe}}$ = \SI{3e14}{\watt/\centi\meter\squared}) with respect to the kick pulse ($\lambda = 800~\textup{nm}$, $\tau_\text{FWHM} = 200~\textup{fs}$, I$_{\textup{kick}}$ = \SI{2e13}{\watt/\centi\meter\squared}) is given by the numbers on the top of the vertical panels.
     % As expected the F$^+$ ions, whose phenyl plane has a large angle with respect to the SMPA,
     % split into four easily resolvable regions 0.67 ps after the kick pulse. In the following
     % time delays approaching 4 ps this split is seen to oscillate between small and large angles.
     %Investigation of the Br$^+$ ion images at the same time delays, does not directly reveal a split
     % into the expected four region structure since the angle of the phenyl plane with bromine
     % substituents will reside close to the SMPA, and these will therefore tend to merge.
     %Examination of both ion images and the corresponding the angular distributions, however, reveal minima in the ion yield detected along the SMPA, and oscillation of the two phenyl-rings.
     %Extracting the angles of the two phenyl planes with respect to the SMPA it is possible to directly monitor the evolution of the dihedral angle of the molecule -- See text for details.
     }
  \label{fig:Bp_ion_evo_short}
\end{figure*}

To investigate the torsional motion induced by the kick pulse we recorded series of both F$^+$ and
Br$^+$ images at different kick-probe delays.
The probe pulse was polarized perpendicular to the detector plane to ensure a detection efficiency that is independent of the orientation of the F- and Br-phenyl-rings. The results are shown in panels A and B of \autoref{fig:Bp_ion_evo_short}.
The first striking observation is that from 0.67 ps to 4.0 ps the F$^+$ images develop a pronounced 4-peak structure consisting of a pair in the top and a pair in the bottom part of the images. This shows that the kick pulse sharpens the alignment of the SMPA in agreement with a recent study on naphthalene where it was demonstrated that the combined action of 3D adiabatic prealignment and a short kick pulse could significantly improve the alignment of the molecular plane in a time interval shortly after the kick pulse \cite{Hansen:PRL:2011}. The second observation is that the angular separation of the two regions in each F$^+$ pair oscillates as a function of time. At $t$ = 1.33 ps the angular separation reaches a local minimum, at $t$= 2.0 ps a local maximum, at $t$ = 2.67 ps a local minimum etc.  Turning to the Br$^+$ ion images it is seen that the confinement along the Y-axis (vertical) improves. Unlike the F$^+$ images a clear 4-peak structure is not observed. We ascribe this to the fact that the Br-phenyl ring is much closer to the SMPA, i.e., Br$^+$ ions from molecules with the Br-phenyl ring localized on each side of the SMPA will overlap and prevent a distinct 4-peak structure. It is, however, observed that the width of the Br$^+$ oscillates as a function of time in phase with that observed in the F$^+$ images. These joint observations show that the dihedral angle between the Br-phenyl ring and the F-phenyl ring changes as a function of time. In other words, the torsional motion is directly imprinted on the ion distributions recorded.

The contrast can be improved significantly by recording the F$^+$ and Br$^+$ images with the probe pulse polarized parallel to the kick pulse polarization.
%Before proceeding to the quantitative analysis of the ion images we note that the F$^+$ and Br$^+$ images were also recorded with the probe pulse polarized parallel to the kick pulse polarization.
These data are displayed in \autoref{fig:Bp_ion_evo_short}, panels (c) and (d). The evolution of both ion species is the same as that observed in rows (a) and (b), but the contrast is significantly higher. In particular, the 4-peak structure in the F$^+$ images is very distinct with complete separation between the two regions in each pair and the oscillations in the angular width of the Br$^+$ images stand out very clearly.
Our qualitative understanding of the improved contrast is that the probe pulse preferentially ionizes molecules with their SMPA close to its polarization axis.
%{\bf just like a linear molecule is usually most efficiently ionized, by a strong field, if the field is parallel to the internuclear axis DISCUSS WITH HS}.
In other words, the probe pulse selects the molecules with the strongest alignment of the SMPA.

The quantitative analysis is performed by fitting the angular distributions, extracted from the ion
images, with a sum of four Gaussian functions. From these fits, the average angles, $\langle\phi_{\textrm{Br}^{+}} \rangle$,
of the Br-phenyl ring, with respect to the kick pulse polarization, can be extracted directly, from the peak
positions of the Gaussian functions. Similarly, the average angle, $\langle\phi_{\textrm{F}^{+}} \rangle$, of the
F-phenyl-ring can be found from the fit of the F$^{+}$ angular distributions.

\begin{figure}
    \centering
    \includegraphics[width=0.50\textwidth]{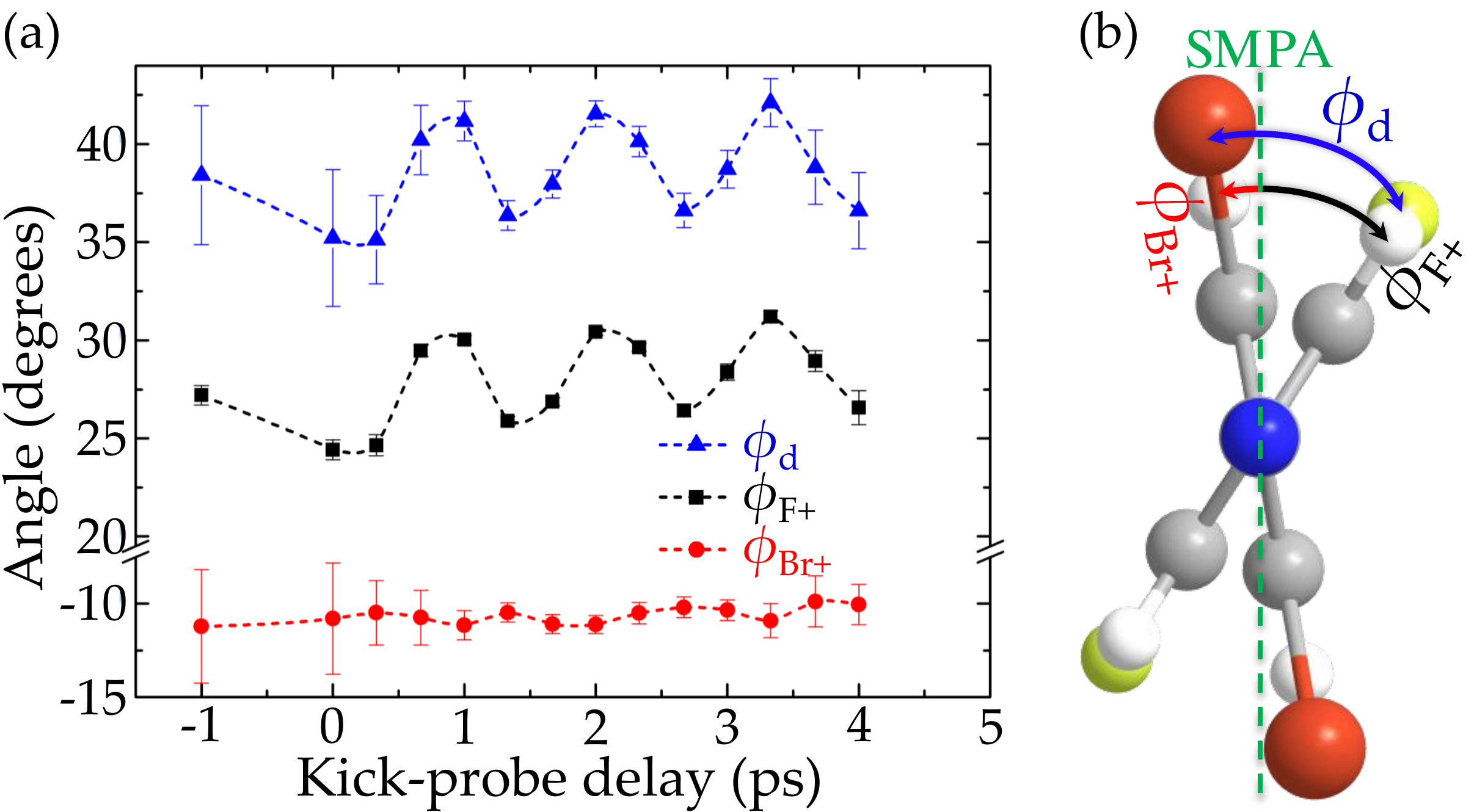}
    \caption{(a) Time-dependence of the dihedral angle (blue triangles) following the kick-pulse, obtained by adding the orientation of the F-phenyl-ring (black squares) and the orientation of the Br-phenyl-ring with respect to the SMPA -- see text for details.
      (b) Sketch of the molecular structure as seen in end view. The SMPA of the molecule and the extracted angles are marked on the drawing.}
  \label{fig:Bp_angle_evo_short}
\end{figure}

Figure \ref{fig:Bp_angle_evo_short} displays the average angle between the F-phenyl-rings (black squares) and the kick pulse polarization as a function of kick-probe time delay for the time interval where a clear four-peak-structure can be identified in the angular distributions.
The dashed black line splined to the experimental measurements indicates that the oscillations observed in the images of the F-phenyl ring have an amplitude of $\sim$2.5\degree with respect to the kick pulse polarization direction and a period of $1.25$~ps. Closer examination of the oscillating trace reveals a small overall increase towards larger $\langle\phi_{\textrm{F}^{+}} \rangle$ which shows that the plane of the F-phenyl-ring slowly moves away from the kick pulse polarization. Also the average angle between the Br-phenyl-rings (red circles) and the kick pulse polarization is shown. In the time interval 1--4~ps the uncertainties on the measurements are small and allows for determination of the period (1.25~ps) and amplitude ($\sim$0.5\degree) of the motion. For the Br-phenyl-ring, we observe a small decrease in the $\langle\phi_{\textrm{Br}^{+}} \rangle$ with increasing time delay. The slight decrease in $\langle\phi_{\textrm{Br}^{+}} \rangle$ and corresponding increase in $\langle\phi_{\textrm{F}^{+}} \rangle$, are fully compatible with an overall rotation of the molecule about the stereogenic axis.

The average values, $\langle\phi_{\textrm{F}^{+}}\rangle$ and $\langle\phi_{\textrm{Br}^{+}} \rangle$, can be used to
calculate the instantaneous dihedral angle, $\phi_{\textrm{d}}$, of the molecule by simple addition (see \autoref{fig:Bp_angle_evo_short} (b)). By retrieving these average values at every kick-probe time
delay it is possible to directly monitor the evolution of the dihedral angle of the molecule.  The results for the dihedral angle ($\phi_{\textrm{d}}$, blue triangles) are displayed in \autoref{fig:Bp_angle_evo_short} (a), with a dashed line included to help guide the eye and emphasize the torsional motion. The oscillation is seen to have an amplitude of $3\degree$ and a period of 1.25~ps.

\begin{figure}
    \centering
   \includegraphics[width=0.45\textwidth]{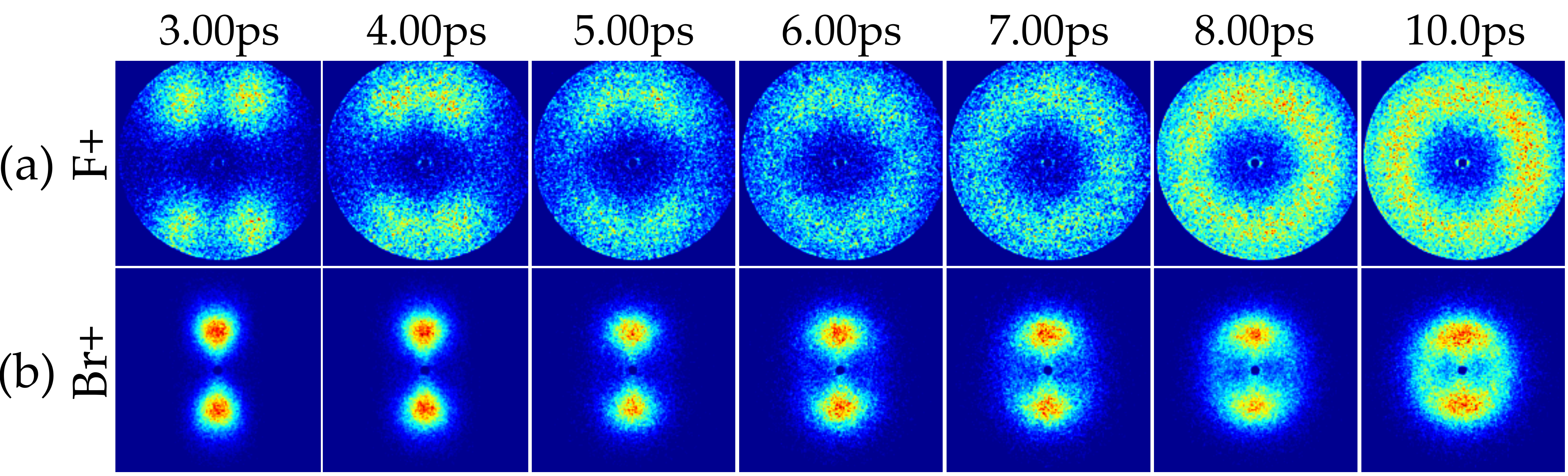}
   \caption{F$^+$ (a) and Br$^+$ (b) ion images obtained at long ($t \geq 3$~ps)
     kick-probe time delays obtained using a probe pulse linearly polarized perpendicular to the detector.
     The delay of the probe pulse ($\lambda = 800~\textup{nm}$, $\tau_\text{FWHM} = 30~\textup{fs}$, I$_{\textup{probe}}$ = \SI{3e14}{\watt/\centi\meter\squared}) with respect to the kick pulse ($\lambda = 800~\textup{nm}$, $\tau_\text{FWHM} = 200~\textup{fs}$, I$_{\textup{kick}}$ = \SI{2e13}{\watt/\centi\meter\squared}) is given by the numbers on the top of the vertical panels.}
  \label{fig:Bp_ion_evo_long}
\end{figure}

We also recorded F$^+$ and Br$^+$ ion images for times larger than 4 ps. The results, including the 3.0 ps data, are shown in \autoref{fig:Bp_ion_evo_long}.
For $t >$ 4.0 ps the F$^+$ ion images loose their distinct four region structure and, instead, evolve into a circular shape at $t \geq 8$~ps. This prevents tracking the evolution of $\langle\phi_{\textrm{F}^{+}}\rangle$. The images, however, confirm that the stereogenic axis of the molecule is still tightly confined along the major axis of the YAG polarization ellipse, since the central region is absent from ions.
The corresponding Br$^+$ ion images show a similar delocalization behavior, but the broadening is slower. This observation is consistent with theory as discussed in \autoref{sec:dephasing}.

\begin{figure}
    \centering
    \includegraphics[width=0.45\textwidth]{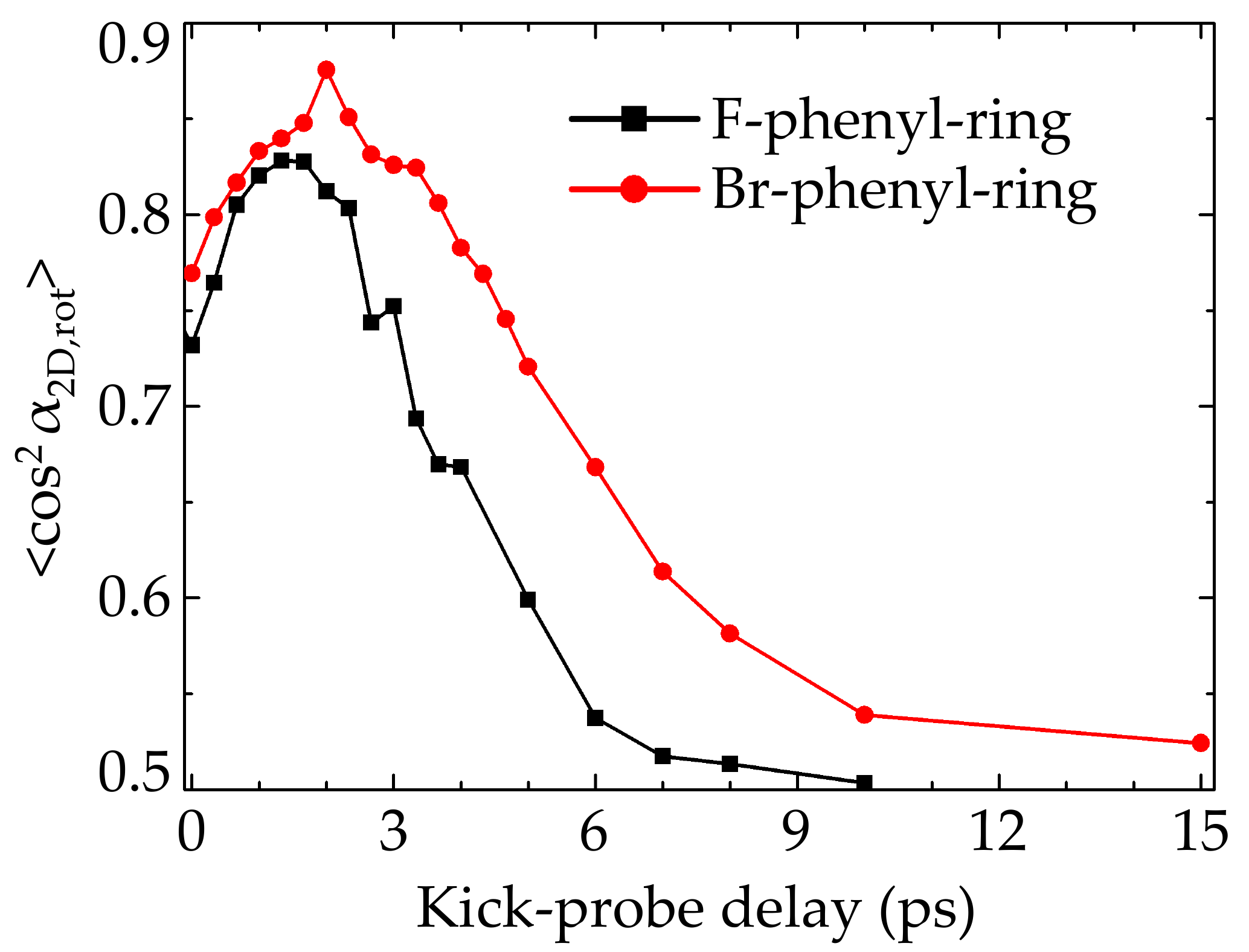}
    \caption{Time dependence of the planar alignment of the individual phenyl-rings. The $\langle \cos^2\alpha_{2D} \rangle$ values are calculated from rotated angular distributions to remove the effect of the torsion (see text for details). Black squares (red circles) correspond to the confinement of the F-phenyl-ring (Br-phenyl-ring).}
  \label{fig:Dephasing}
\end{figure}

To quantify the time-dependence of the delocalization in the F$^+$ and Br$^+$ ion images we determined for each image $\langle \cos^2\alpha_{2D} \rangle$, where $\alpha_{2D}$ is the angle between the projection of an ion velocity vector on the detector plane and the Y-axis (See \autoref{fig:Autovar1}).
%Our analysis aims at identifying the contribution of rotation to the delocalization and therefore we must eliminate the effect of torsion.
This is implemented by rotating the peaks fitted to the angular distributions of F$^+$ and Br$^+$ ions, towards the Y-axis by the following procedure:
%To investigate the dephasing dynamics of the different rings in greater detail, the planar alignment was extracted for each of the phenyl planes at all kick-probe time delays. To get the most correct picture of the dynamics, the peaks fitted to the angular distributions extracted from the fluorine and bromine images were rotated towards the Y-axis (see \autoref{fig:Schematic_setup}) by the following routine:
If the peak is located in the first or third quadrant the F$^+$ (Br$^+$) image is rotated by $-\langle\phi_{\textrm{F}^{+}}\rangle$ ($-\langle\phi_{\textrm{Br}^{+}}\rangle$) before calculation of $\langle \cos^2\alpha_{2D} \rangle$, whereas if the peak is in the second or fourth quadrant the image is rotated by $+\langle\phi_{\textrm{F}^{+}}\rangle$ ($+\langle\phi_{\textrm{Br}^{+}}\rangle$).
This procedure allows for calculation of the best estimate of the spread from the mean by including all data.
 %aids to combine the split distributions observed in the images (and angular distributions) along the Y-axis and thereby give the most correct value for the planar confinement.
The $\langle \cos^2\alpha_{2D} \rangle$ values are displayed in \autoref{fig:Dephasing}.
A clear difference in the delocalization dynamics of the two phenyl planes is observed, with the light fluorine substituted phenyl ring approaching $\langle \cos^2\alpha_{2D} \rangle$ = 0.5 more rapidly than the heavier bromine substituted phenyl ring. This is fully consistent with semi-classical theoretical predictions presented in \autoref{sec:dephasing}.
%Here we see a noticeable difference in the dynamics by which the two phenyl rings change the spread of their mean value. This is clearly visible by inspecting the times where $\langle \cos^2\alpha_{2D} \rangle$ approaches 0.5, which occurs for the F-phenyl-ring between 8-10 ps and for the Br-phenyl-ring at $t >$15 ps.
%It should be noted that the same quantitative behavior is observed when extracting the planar alignment from the raw data, i.e. unrotated, and hence these lead to the same conclusions. However, for the unrotated data the alignment trace is overlaid by the additional torsion dynamics which produces slightly higher (lower) $\langle \cos^2\alpha_{2D} \rangle$ values when $|\langle\phi_{\textrm{F,Br}^{+}\textrm{-kick. pol}}\rangle |$ is small (large).}
%{\bf does this make sense?! If we assume that the spread is the same for the two rings I believe we would still observe a difference in the $\langle \cos^2\alpha_{2D} \rangle$ since the starting points of the Br and F ring are different - Two rings which are located far apart will of course spread faster than two rings lying close (on top of each other) - The Br ring will therefore have "an advantage" which could account for at least some of the observed difference.}

\subsection{Covariance and Autovariance Mapping}
\label{Autovariance}

Before discussing the theoretical results we introduce covariance mapping analysis \cite{Frasinski:science:1989} to the experimental data. This analysis strengthen our interpretation of torsional motion, based on the fluorine and bromine ion images in \autoref{fig:Bp_ion_evo_short}
%, we have implemented a modified version of covariance mapping \cite{Frasinski:science:1989} to the experimental data.
Covariance mapping is a technique to reveal correlations, which in the high count rate regime would
otherwise become blurred or completely lost due to the large contributions from uncorrelated events.
The basic principle is to calculate the covariance, i.e., to obtain a crosscorrelation of the variance in the data \cite{goodman:1985},
%of the technique is to obtain an autocorrelation of the variance in the data;
to extract the deviation from the mean value for all the individual frames of the ion image, pertaining to each (probe) laser shot. This is fundamentally different from coincidence measurements in, e.g., COLTRIMS and reaction-microscopes \cite{Dorner:PhysRep:2000, Ullrich:RopiP:2003} where the count rate is
restricted to less than one event per laser shot, since the covariance map is produced from the
correlations of the variance, which per definition has no restrictions on the number of events
obtained per laser shot.

The covariance ($C \left( x,y \right)$) between two observables $X(x)$ and $Y(y)$ is
defined as the mean of the product between the observables $\langle X(x)Y(y) \rangle$ subtracted by the product of the means $\langle X(x)\rangle \langle Y(y) \rangle$:
\begin{equation}
  \begin{split}
    C(x,y) =& \mean{(X(x) - \mean{X(x)})(Y(y) - \mean{Y(y)})} \\
%    =& \mean{XY - X\mean{Y} - Y\mean{X} + \mean{X}\mean{Y}}\\
%    =& \mean{XY} - \mean{X}\mean{Y} - \mean{X}\mean{Y} + \mean{X}\mean{Y}\\
%    =& \mean{XY} - \mean{X}\mean{Y}
    =& \mean{X(x)Y(y)} - \mean{X(x)}\mean{Y(y)}\\
%    =& \frac{1}{N} \sum^{N}_{i=1}{X_{i}(x)Y_{i}(y)} - \left[ \frac{1}{N} \sum^{N}_{i=1}{X_{i}(x)}
%    \right] \left[ \frac{1}{N} \sum^{N}_{j=1}{Y_{j}(y)} \right]
  \end{split}
  \label{eq:Covar1}
\end{equation}
Here $X$ and $Y$ correspond to the observed signals, with $x$ and $y$ being the variables
investigated for correlation.

\begin{figure}
  \centering
   \includegraphics[width=0.47\textwidth]{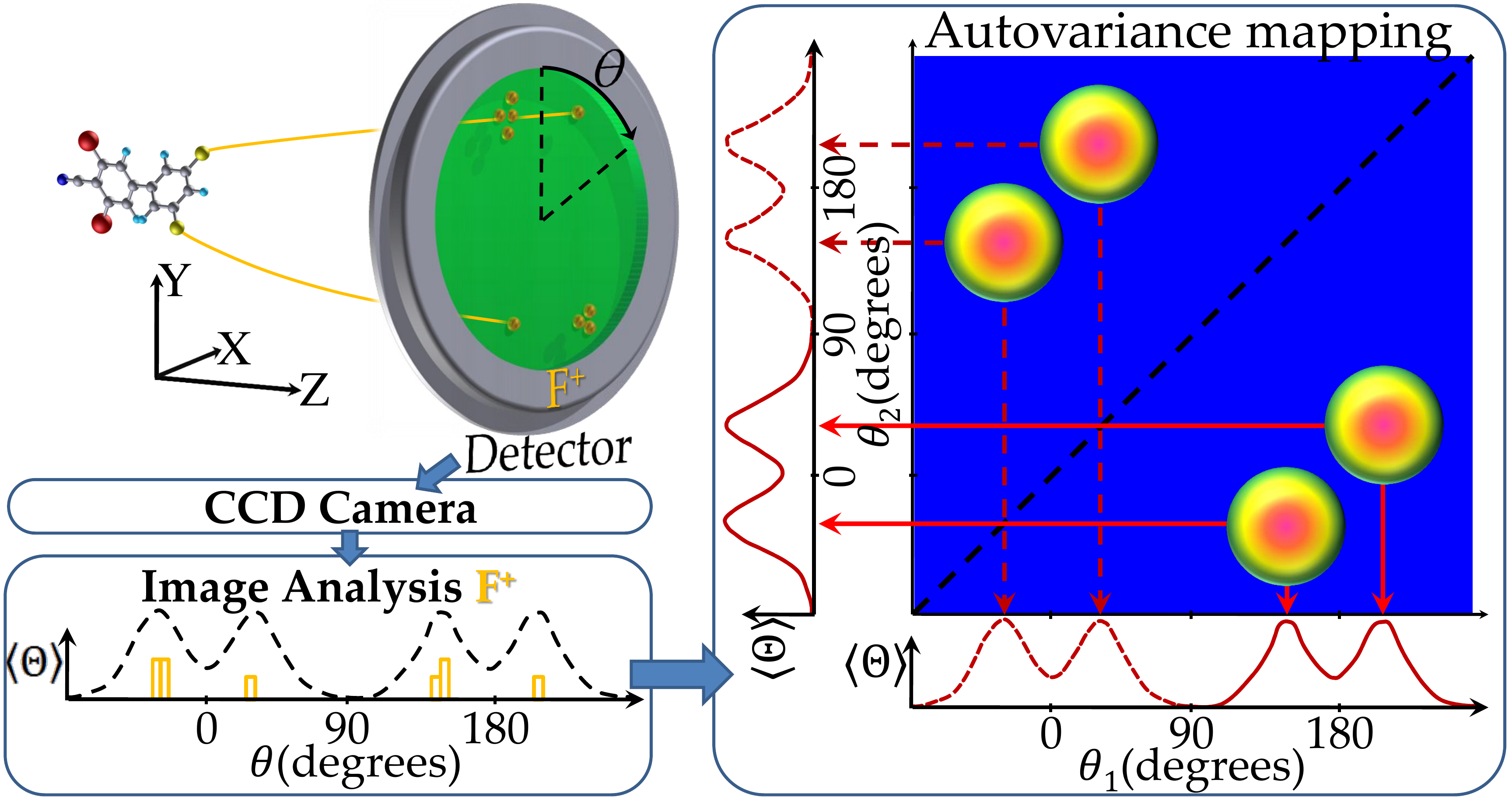}
   \caption{The autovariance mapping principle illustrated for F$^+$ fragments produced by coulomb
     explosion of laser aligned DFDBrCNBph. From the autovariance map seen on the right it is possible
     to extract correlation information between the ejection angle of the F$^+$ ions.
     The blue arrows indicate the steps in the data acquisition and data processing leading to the production of autovariance maps.
     %Autovariance maps for different series of mass/charge ratios can easily be generated simply by recording images at a specified time delay determined by a negative HV fast pulser gate connected to the front of the MCP.
     }
  \label{fig:Autovar1}
\end{figure}

Although many different correlation techniques have been explored previously and found many different applications within natural science \cite{brown:nature:1956,Aue:JCP:1976,Ullrich:RopiP:2003,schroter:science:2011}, to our knowledge no reports on using covariance mapping to extract angular correlation information from 2D ion images has been reported.
%As shown in \autoref{eq:Covar2}
We suggest to do this by replacing $X$ and $Y$ with $\Theta$ and $\Xi$,
corresponding to the angular distributions of the ion signal, and $x$ and $y$ with $\theta$ and
$\xi$ being the ion detection angle with respect to the Y-axis (\autoref{fig:Schematic_setup}).
%\begin{equation}
%  \begin{split}
%    C(\theta,\phi) =& \mean{(\Theta(\theta) - \mean{\Theta}) (\Phi(\phi) - \mean{\Phi})} \\
%    =& \mean{\Theta\Phi}-\mean{\Theta}\mean{\Phi}=
%    \mean{\Theta(\theta)\Phi(\phi)}-\mean{\Theta(\theta)}\mean{\Phi(\phi))} \\
%    =& \frac{1}{N} \sum^{N}_{i=1}{\Theta_{i}(\theta)\Phi_{i}(\phi)} - \left[ \frac{1}{N}
%      \sum^{N}_{i=1}{\Theta_{i}(\theta)} \right] \left[\frac{1}{N} \sum^{N}_{j=1}{\Phi_{j}(\phi)} \right]
%    \label{eq:Covar2}
%  \end{split}
%\end{equation}
%Applying \autoref{eq:Covar2}
Applying this version of covariance mapping to the impact coordinates of the ions in the 2D velocity map images, allows for extraction of the angular correlation between ejected fragments.
In our experiment only one ion species, e.g., F$^+$, is detected per laser shot and thus $\Theta = \Xi$.
As a consequence Eq. \eqref{eq:Covar1} may be written as
%However, if only one detector is used, as is the case here, $\Theta$ will be equal to $\Xi$, in which case the expression for correlation becomes
\begin{equation}
  \begin{split}
    %C(\theta_1,\theta_2) =& \mean{(\Theta(\theta_1) - \mean{\Theta(\theta_1)}) ((\Theta(\theta_2) - \mean{\Theta(\theta_2)})} \\
C(\theta_1,\theta_2)
    =& \mean{\Theta(\theta_1)\Theta(\theta_2)}-\mean{\Theta(\theta_1)}\mean{\Theta(\theta_2)}\\
%    =  \mean{\Theta(\theta)\Theta(\theta)}-\mean{\Theta(\theta)}\mean{\Theta(\theta))} \\
    =& \frac{1}{N} \sum^{N}_{i=1}{\Theta_{i}(\theta_1)\Theta_{i}(\theta_2)} \\
    -& \left[ \frac{1}{N}
      \sum^{N}_{i=1}{\Theta_{i}(\theta_1)} \right] \left[\frac{1}{N} \sum^{N}_{i=j}{\Theta_{j}(\theta_2)} \right]
    \label{eq:Covar3}
  \end{split}
\end{equation}
Since this corresponds to investigating correlations of a signal with itself this approach is equivalent to obtaining an autocorrelation of the variance in the data and so the 2D maps depicting the correlation should be termed autovariance maps.
%The limitation of using only one detector restricts this technique to investigate angular correlations between ions having identical mass/charge ratios, which is only meaningful when the molecule contains two or more identical observables. Since the correlation of a signal with itself corresponds to an autocorrelation, \autoref{eq:Covar3} is also known as the autovariance function {\bf [ref. Goodman Statistical Optics Wiley 1985 ?!]}, and the single detector setup can be used to obtain autovariance maps.
The principle of autovariance mapping is illustrated in \autoref{fig:Autovar1}.
Following each laser shot the ions focused onto the detector are recorded by a CCD camera after which an image analysis program saves the impact coordinates, illustrated by the yellow bars in the image analysis box.
The experiment is repeated $N$ times, typically $N = 10.000$, giving the averaged ion images (such as those shown in \autoref{fig:Bp_ion_evo_short}) with the averaged angular distributions (\autoref{fig:Autovar1} dashed black line). The autovariance $C(\theta_1,\theta_2)$ can now be determined by applying Eq. \eqref{eq:Covar3} to all $N$ frames of the ion image, and displayed as a 2D map of the ejection angles.
%To overcome frame-to-frame fluctuations in the number of ions detected and their positions the data is averaged over thousands of experimental realizations (dashed black line).
%From such a data set the correlation between the detected ions in the individual frames ($C \left( \theta_1,\theta_2 \right)$) can be displayed as a 2D map of their ejection angles.
In this map the two axes correspond to the angular distribution from the averaged image, with the positive Y-axis denoting 0$\degree$.
% (see \autoref{fig:Autovar1}).

Autovariance maps obtained through this procedure will tend to produce a strong autocorrelation line along the diagonal of the map, since the observation of an ion in frame $\Theta_i$ at the angle $\theta_1$ is naturally also observed in the same frame, for the identical observable $\Theta_i$, at $\theta_2=\theta_1$.
This autocorrelation trace will also act as an axis of symmetry for the correlation, since $C(\theta_1,\theta_2) = C(\theta_2,\theta_1)$, which is in fact equivalent to the square of the r.m.s. standard deviation by definition.
The positive areas of the autovariance maps, which fall outside the autocorrelation line $(\theta_1 \neq \theta_2)$, indicate the correlation between ions at $\theta_1$ and $\theta_2$, showing an increased likelihood of observing an ion at $\theta_2$ provided that an ion is detected at $\theta_1$.

\begin{figure}
   \centering
   \includegraphics[width=0.47\textwidth]{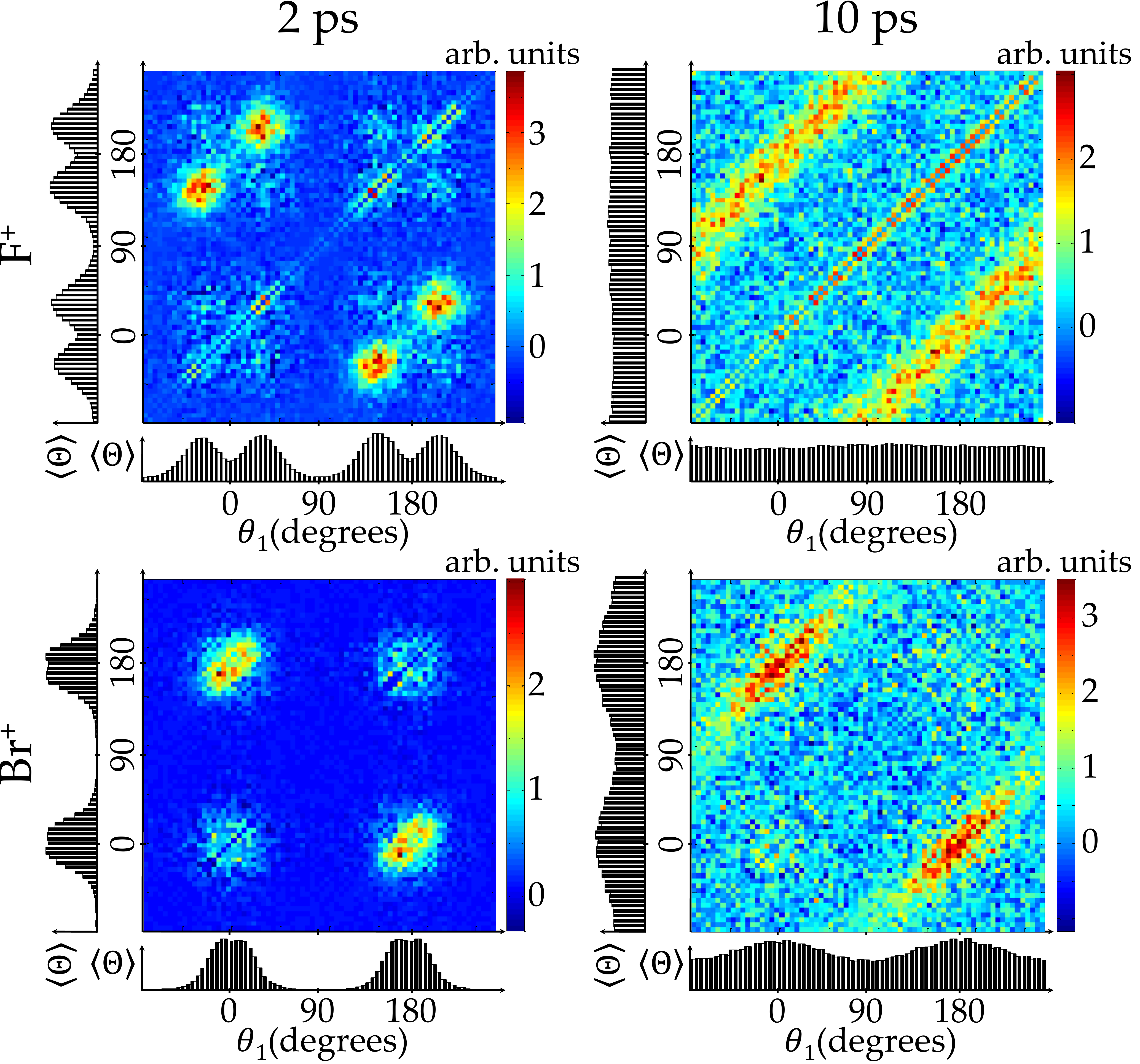}
   \caption{Autovariance maps for F$^+$ (upper panels) and Br$^+$ ions (lower panels) at kick-probe time delays of 2 ps (left column)  and 10 ps (right column).
   To smoothen out small fluctuations, the data has been binned in steps of 5 degrees. Also the large signal, due to autocorrelation, along the diagonal of the maps, has been set to zero to reveal the less intense correlations in the autovariance map.}
  \label{fig:Autovar2}
\end{figure}

Figure~\ref{fig:Autovar2} shows the angular autovariance maps obtained from both F$^+$ and
Br$^+$ ion images at two different kick-probe time delays. In these maps the dominant autocorrelation signal along the diagonal has been set to zero to increase the contrast of the interesting but less intense correlations in the autovariance map. Also these data sets have been binned in steps of 5$\degree$ to smoothen out small fluctuations.
A striking feature of all autovariance maps is that the observed correlations are restricted to areas where the angle between the ejected ions is close to $180\degree$. This is seen, e.g., in the F$^+$ map at 2 ps, where the signal at $\sim$150\degree ($\sim$210\degree) is correlated to the region at $\sim$-30\degree ($\sim$30\degree).
In other words, when an F$^+$ ion is ejected upwards at an angle of $150\degree$ there is a high probability that another F$^+$ ion is ejected at -30\degree , i.e., just in the opposite direction.
This corroborates our interpretation of the two ion species as direct observables of the orientation of each of the phenyl rings.
In addition, the observation of four prominent positive areas in the autovariance map at $t$ = 2 ps confirms the interpretation of the 4-peak structure in the F$^+$ ion images (\autoref{fig:Bp_ion_evo_short}).
Similar considerations hold for the Br$^+$ autovariance map at $t$ = 2 ps. Notably the appearance of a 4-peak structure with positive covariance shows even clearer than the averaged ion image on \autoref{fig:Bp_ion_evo_short}, that the Br-phenyl plane has a small angular offset with respect to the kick pulse polarization.
At $t$ = 10 ps there is no longer a 4-peak structure but the distinct oblique lines show unambiguously that emission of both F$^+$ and Br$^+$ ions still occurs in a pairs with an upward and a downward ion.
The extension of the covariance principle to obtain angular correlation between ions
allows for the extraction of additional information from the 2D images obtained using VMI spectrometers.
As shown here, a single detector setup, can be used to obtain autovariance maps between ions of identical mass-to-charge ratios, which can be used to both substantiate and clarify the interpretation of the recorded ion images.
This can be extended to obtain the correlation between ions with different mass-to-charge ratios by pulsing the front of the MCP twice per laser shot and running the CCD camera in burst mode.
In addition, we note that the technique can also be extended to obtain radial autovariance maps, for fragmentation processes with multiple energy channels, or combined to encompass both angular and radial correlation of the 2D ion images.

\section{Comparison with theory}
\label{theory}

The main findings of \autoref{Experimental Results} were that the dihedral angle oscillates with a period of approximately 1.25 ps, and an amplitude  of 3$\degree$. These findings were based on an analysis of the data assuming that the two ion species move out to the detector in the plane of the (substituted) phenyl ring from which they originated. The covariance and autovariance analysis introduced validated this assumption and we may build the theory on this fact.

\subsection{Early time dynamics: Torsional motion}
The purpose of this section is to provide a theoretical foundation for understanding the experimentally observed ion images shown in \autoref{fig:Bp_ion_evo_short}.
%These ion images essentially show the distributions of the angles of the F-phenyl-ring, $\phi_{\text{F}^+}$, and the Br-phenyl-ring, $\phi_{\text{Br}^+}$.
Atomic units $(|e|=\hbar=m_\text{e}=a_0=1)$ are used throughout.
A useful starting point is to discuss the widths of the ion images. These may conveniently be quantified in terms of the variance
\begin{equation}
\sigma_\text{i}^2=\langle \phi_\text{i}^2\rangle-\langle\phi_\text{i}\rangle^2, \quad \text{i}={\text{Br}^+},{\text{F}^+}.
\end{equation}

For the sake of interpretation, we rewrite these two angles in terms of the dihedral angle, $\phi_\text{d}$, and the overall rotation, $\Phi$ [see \autoref{fig:Bp_angle_evo_short} (b)]:
\begin{align}
\phi_{\text{Br}^+}&=\Phi+\eta\phi_\text{d},\\
\phi_{\text{F}^+}&=\Phi-(1-\eta)\phi_\text{d},\\
\eta&=\frac{I_{\text{F}^+}}{I_{\text{F}^+}+I_{\text{Br}^+}}.
\end{align}
Consequently, we may express the variance
%, $\sigma_\text{i}^2=\langle \phi_\text{i}^2\rangle-\langle\phi_\text{i}\rangle^2$,
of the observed ions as
\begin{align}
\sigma_{\phi_{\text{Br}^+}}^2&=\sigma_{\Phi}^2+\eta^2\sigma_{\phi_\text{d}}^2\nonumber\\
&+2\eta(\langle\Phi\phi_\text{d}\rangle-\langle\Phi\rangle\langle\phi_\text{d}\rangle),\label{eq:sigmabr}\\
\sigma_{\phi_{\text{F}^+}}^2&=\sigma_{\Phi}^2+(1-\eta)^2\sigma_{\phi_\text{d}}^2\nonumber\\
&-2(1-\eta)(\langle\Phi\phi_\text{d}\rangle-\langle\Phi\rangle\langle\phi_\text{d}\rangle).\label{eq:sigmaf}
\end{align}

Previous investigations show that the variance of the dihedral angle is small (See Figs.~3 and~9 in Ref.~\cite{Madsen:JCP:2009}). It is therefore obvious to ask if  $\sigma_\Phi^2\gg \sigma_{\phi_\text{d}}^2$. As follows from Eqs.~\eqref{eq:sigmabr} and~\eqref{eq:sigmaf} this can only be the case if the widths of the ion images are comparable. Inspection of Figs.~\ref{fig:Bp_ion_evo_long} and~\ref{fig:Dephasing} confirms that for long times ($t > 3$ ps) this is not true. For shorter times, however, the widths of the two ion images are close (see \autoref{fig:Bp_ion_evo_short}). Further, we have previously argued for separability of the $\Phi$ and $\phi_\text{d}$ coordinates in this limit \cite{Madsen:JCP:2009} meaning that the last term in Eqs.~\eqref{eq:sigmabr} and~\eqref{eq:sigmaf} vanishes completely. In this regime $\Phi$ is an adiabatic parameter and we simply discuss the short-term dynamics based on calculations with $\Phi$ fixed meaning that only torsion takes place.

The torsional motion is then dictated by the time-dependent Schr\"{o}dinger equation
\begin{equation}
i\partial_t\Psi(\Phi;\phi_\text{d},t)=\left[T_\text{d}+V_\text{tor}(\phi_\text{d})+V_\text{kick}(\Phi;\phi_\text{d},t)\right]\Psi(\Phi;\phi_\text{d},t),
\end{equation}
with
\begin{equation}
T_\text{d}=-\frac{I_{\text{Br}^+}+I_{\text{F}^+}}{2I_{\text{Br}^+}I_{\text{F}^+}}\frac{\partial^2}{\partial\phi_\text{d}^2}
\end{equation}
being the rotational kinetic energy due to torsion, $V_\text{tor}$ the torsional potential and $V_\text{kick}$ the polarizability interaction energy between the kick pulse and the molecule:
\begin{align}
V_\text{kick}(\Phi,\phi_\text{d},t)&=-\frac{1}{4}F_0^2(t)[\alpha_\text{xx}(\phi_d)\cos^2(\Phi+\eta\phi_\text{d})\nonumber\\
&+\alpha_\text{yy}(\phi_\text{d})\sin^2(\Phi+\eta\phi_\text{d})\nonumber\\
&-2\alpha_\text{xy}(\phi_\text{d})\cos(\Phi+\eta\phi_\text{d})\sin(\Phi+\eta\phi_\text{d})].
\end{align}
We solve this equation for DFDBrBPh using a close coupling method with a value of $\Phi$ corresponding to the SMPA and kick pulse polarization aligned. As initial state $\Psi(t_0)$, we use a state localized in the torsional well at $39^\circ$. The details have been given in Ref.~\cite{Madsen:JCP:2009}.

\begin{figure}
    \centering
    \includegraphics[width=0.45\textwidth]{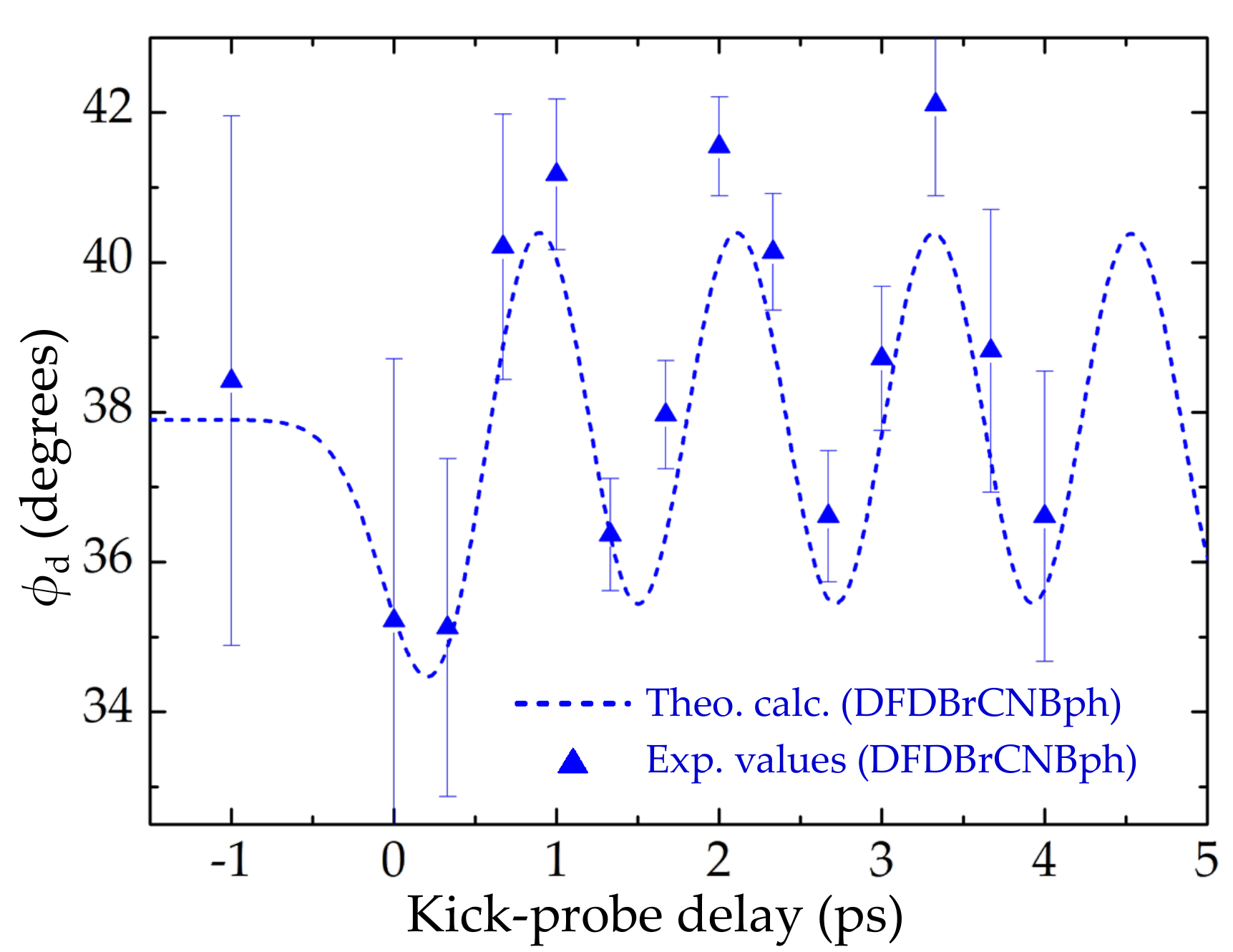}
    \caption{Comparison: Blue triangles - experimental results, dashed blue line - theoretical
      calculation on DFDBrBPh shifted by $-0.9\degree$ to match the dihedral angle of DFDBrCNBPh at equilibrium.}
  \label{fig:Bp_angle_evo_short_2}
\end{figure}

Figure \ref{fig:Bp_angle_evo_short_2} shows the result of the calculation. The torsional dynamics consist of small, periodic oscillations. As argued previously \cite{Madsen:JCP:2009}, these oscillations stem from the fact that the kick pulse leaves DFDBrBPh in a coherent superposition of the ground state and first excited state of the torsional potential --- essentially simple harmonic oscillator states separated by 3.42 meV.
%De 3.42 meV kommer fra JCP artiklen: i Caption Fig. 3 p. 3. Her står der at afstanden fra minimum i potentialbrøden til nulpunktsvibrationen er 1.71 meV. Denne er givet som E_0 = hv(0+½)=½hv => Afstanden mellem energiniveauerne i den harmoniske oscillator er DeltaE = E_(n) - E_(n-1) = hv = 3.42 meV.
The molecule used in the experiment is DFDBrCNBPh rather than DFDBrBPh.
The torsional potential differs, however, only slightly from that used previously for DFDBrBPh.
% This implies that the actual torsional potential in the experiment differ from the torsional potential used in the current calculation.
Since the kick pulse only populates the first few torsional states where the potential is very close to harmonic, the main effect to take into account is the shift of the minima of the torsional potential from one molecule to the other ($-0.9^\circ$). It follows that we may simply shift the previously calculated theoretical curve by $-0.9^\circ$ to obtain a comparison with the experimental results. As seen in \autoref{fig:Bp_angle_evo_short_2} this leads to an impressive agreement with the experimental results.

\subsection{Long time behavior: Dephasing of the phenyl rings}
\label{sec:dephasing}
For longer time scales ($t>3$ ps), the semi-classical model is no longer attractive. The  reasons are that  the widths of the two ion images (Figs.~\ref{fig:Bp_ion_evo_short} and~\ref{fig:Bp_ion_evo_long}) start to deviate from one another meaning that the variance is no longer dominated by $\Phi$ alone and, hence,  that the separability in $\Phi$ and $\phi_\text{d}$ breaks down \cite{Madsen:JCP:2009, Coudert:PRL:2011}.

To provide a qualitative analysis in  this long-time regime, we exploit a different theoretical limit, where the phenyl-rings are treated as rigid rotors. In this case, each of the rings $\text{i}=\text{Br}^+,\text{F}^+$ can be described by a wave packet
\begin{equation}
    \Psi_\text{i}(\phi_\text{i},t)= \sum_{J_\text{i}}c_{J_\text{i}}\frac{1}{\sqrt{2\pi}}e^{iJ_\text{i}\phi_\text{i}}e^{-i\frac{J_\text{i}^2}{2I_\text{i}}t}
    \label{eq_ring_wavepacket},
\end{equation}
where $J_\text{i}$ labels a rotational state and where $c_{J_\text{i}}$ is a fixed constant after the end of kick pulse.

To estimate on what time scale, $\tau_\text{i}$, such a wave packet dephases, we look at the time evolution of the coherence between two neighboring levels $J_\text{i}$ and $J_\text{i}+1$.
It immediately follows from \eqref{eq_ring_wavepacket} that this coherence beats like
%\begin{align}
%\vert \Psi_\text{i}(\phi_\text{i},t)\vert^2\simeq\vert &= \vert c_{J_\text{i}}\vert^2+\vert c_{J_\text{i}+1}\vert^2\nonumber\\
%&+2\text{Re}(c_{J_\text{i}}c_{J_\text{i}+1}) \cos\left(\frac{J_\text{i}+1}{2I_\text{i}}t\right)
%\end{align}
\begin{equation}
\cos\left(\frac{2 J_\text{i}+1}{2I_\text{i}}t\right)
\end{equation}
%CBM: JEG FAAR EN FAKTOR 2 PAA J'ET I FORHOLD TIL DIG. JEG HAR SAT DET IND IN (11) OG (12) ER DU ENIG? JEG ER ENIG
Consequently, the two rings dephase at a ratio
\begin{equation}
r=\frac{\tau_{\text{F}^+}}{\tau_{\text{Br}^+}}=\frac{I_{\text{F}^+}}{I_{\text{Br}^+}}\cdot \frac{2J_{\text{Br}^+}+1}{2J_{\text{F}^+}+1}.\label{eq:ratio}
\end{equation}
The value of $J_i$ depends on the rotational kinetic energy, $E_\text{kin,i}$, of each ring. Due to the $a$ times higher polarizability of the Br-phenyl-ring compared to the F-phenyl-ring, we assume that $E_\text{kin,Br}=a \times E_\text{kin,F}$ with $E_\text{kin,i}=J_\text{i}^2/2I_\text{i}$. This result is achieved by integrating the torque of a classic rigid rotor from $0^\circ$ to $90^\circ$  to obtain an energy estimate. We then immediately have
\begin{equation}
\frac{J_{\text{Br}^+}}{J_{\text{F}^+}}=\sqrt{a\frac{I_{\text{Br}^+}}{I_{\text{F}^+}}}
\end{equation}
Using this in Eq.~\eqref{eq:ratio} yields
\begin{equation}
r\simeq\sqrt{a\frac{I_{\text{F}^+}}{I_{\text{Br}^+}}}
\end{equation}
With a value of 0.21 for the ratio of the moments of inertia and  $a=2$ \cite{Note1}
we obtain $r=0.65$ in qualitative agreement with the dephasing ratio of 0.68 obtained from the data presented in \autoref{fig:Dephasing} by comparing the full width at half maximum values of the Gaussian fits to the ion pictures.

\section{Conclusions}
\label{conclusions}
In conclusion we have shown that it is possible to induce torsion in a substituted biphenyl molecule with a nonresonant kick pulse and image the motion in real-time with an intense delayed probe pulse.
The measurements show a distinct torsional motion with an amplitude of $3\degree$ and a period of 1.25 ps for the first 4 ps after the kick pulse, in excellent agreement with results from our theoretical model. At longer times delocalization of the two phenyl rings of the molecule blurs torsion.

An important prerequisite for the experiment is the ability to keep the molecules fixed-in-space during the time-resolved experiment, practically obtained by 3D aligning them in the adiabatic limit. This places both benzene rings of the molecule perpendicular to a 2D imaging detector. Upon Coulomb explosion by the probe laser Br$^+$ ions from one benzene ring imprints an ion image on the detector screen that uniquely identifies the orientation of this ring. Likewise, F$^+$ ion images identify the orientation of the second ring. The difference between the orientation of the two rings determines the dihedral angle, i.e., the normal vibrational coordinate characterizing torsion. Such a procedure would not be possible for samples of randomly oriented molecules.

The paper also showed that it is possible to apply covariance mapping analysis of ion images. This revealed clear correlations between the emission direction of ions strengthening our interpretation of Coulomb explosion as a direct and unique observable of molecular orientation. Covariance and autovariance methods applied to ion imaging have a potential to provide additional information beyond what is normally extracted from averaged quantities.

\section{Acknowledgement}
\label{acknowledgement}

The work was supported by the
Carlsberg Foundation and the Danish Council for Independent Research (Natural Sciences).
CBM acknowledges support from the Chemical Sciences, Geosciences, and Biosciences Division, Office of Basic Energy Sciences, Office of Science, U.S. Department of Energy.
MPJ was supported by a MICINN Juan de la Cierva research grant (JCI-2009-05953), and the Academy of Finland (136079).
LBM was supported by the Lundbeck Foundation, the Danish Research Council (Grant No. 10-085430) and ERC-StG (Project No. 277767-TDMET)

\section*{Appendix}
\label{appendix}

\subsection*{\texorpdfstring{Synthesis of 3,5-difluoro-3´,5´-dibromo-4´-cyanobiphenyl}{Synthesis of
    3,5-difluoro-3,5-dibromo-4-cyanobiphenyl}}
\label{synthesis}

3,5-Difluoro-4´-aminobiphenyl:\cite{Bumagin_TL_2005}
    To a sample vial was added 4-bromoaniline (172.0 mg, 1.0 mmol), 3,5-difluorophenyl boronic acid (189.5 mg, 1.2 mmol), tetrabutylammonium bromide (32.4 mg, 0.1 mmol), potassium carbonate (276.4 mg, 2 mmol) and water (2 mL). To this was added \SI{100}{\micro\liter} of a PdCl$_2$:EDTA:Na$_2$CO$_3$ stock solution prepared by mixing PdCl$_2$ (17.7 mg, 0.1 mmol), EDTA (37.2 mg, 0.1 mmol) and Na$_2$CO$_3$ (21.2 mg, 0.2 mmol) in water (1 mL). The vial which was fittet with a Teflon sealed screwcap and heated to 100$\degree$C for 45 min after which the reaction was cooled to room temperature. The mixture was diluted with Et$_2$O (25 mL) and CH$_2$Cl$_2$ (10 mL). The organic phase was washed with water (2$\times$15 mL), brine (15 mL) and dried using MgSO$_4$. The solvents were removed under reduced pressure and the crude reaction mixture was purified by column chromatography eluting with CH$_2$Cl$_2$ to afford the title compound as a colorless solid (170.5 mg, 83\%). $^1$H NMR (400 MHz, CDCl$_3$) $\delta$ (ppm) 7.37 (d, 2H, J = 8.2 Hz), 7.07-7.00 (m, 2H), 6.74 (d, 2H, J = 8.2 Hz), 6.69 (tt, 1H, J = 8.8, 2.3 Hz), 3.79 (bs, 2H). $^{13}$C NMR (400 MHz, CDCl$_3$) $\delta$ (ppm) 163.6 (dd, J = 245.0, 13.2 Hz), 147.3, 144.8 (t, J = 9.6 Hz), 129.1 (t, J = 2.5 Hz), 128.2, 115.6, 109.1 (dd, J = 18.4, 6.9 Hz), 101.5 (t, J = 25.4 Hz). $^{19}$F NMR (376 MHz , CDCl$_3$) $\delta$ (ppm) -110.8 (t, J = 8.8 Hz). HRMS. C$_12$H$_9$F$_2$N [M+H$^+$] Calculated: 206.0781. Found: 206.0780.

3,5-Difluoro-3´,5´-dibromo-4´-aminobiphenyl:\cite{Tomioka_JACS_2006}
    3,5-Difluoro-4´-aminobiphenyl (165.0 mg, 0.8 mmol) was dissolved in glacial acetic acid (2.5 mL). Bromine ($\SI{82}{\uL}$, 255.7 mg, 1.6 mmol) dissolved in glacial acetic acid (3.8 mL) was slowly added. The reaction was left stirring for 1 hour. Then water (5 mL) and CH$_2$Cl$_2$ (30 mL) was added. Sat. Na$_2$CO$_3$ was added until the reaction mixture became alkaline. The organic phase was washed with sodium thiosulfate (10 mL), water (10 mL), brine (10 mL) and dried using MgSO$_4$. The organic solvent was removed under reduced pressure affording the title compound as a pale brown solid (271 mg, 93\%). This was used without further purification. $^1$H NMR (400 MHz, CDCl3) $\delta$ (ppm). 7.59 (s, 2H), 7.01-6.96 (m, 2H), 6.74 (tt, 1H, J = 8.84, 2.3 Hz) 4.69 (bs, 2H). $^{13}$C NMR (400 MHz, CDCl$_3$) $\delta$ (ppm) 163.6 (dd, J = 246.6, 13.1 Hz), 130.4 (t, J = 2.6 Hz), 130.3, 109.29 (dd, J = 18.6, 7.2 Hz), 109.20, 102.6 (t, J = 25.2 Hz). $^{19}$F NMR (376 MHz , CDCl$_3$) $\delta$ (ppm) -109.8 (t, J =8.5 Hz). GCMS. C$_{12}$H$_7$F$_2$NBr$_2$ Calculated: 362.99 Found: 363 (100), 203 (36), 175 (24), 101 (14).

3,5-difluoro-3´,5´-dibromo-4´-cyanobiphenyl:
    Reaction mixture 1: 3,5-Difluoro-3´,5´-dibromo-4´-aminobiphenyl (270.0 mg, 0.74 mmol) was dissolved in water (1.0 mL) and glacial acetic acid (3.7 mL). Concentrated sulfuric acid (1.0 mL) was added and the reaction mixture was cooled to 10$\degree$C. Sodium nitrite (56.5 mg, 0.82 mmol) dissolved in a minimum of water was slowly added and the reaction was left stirring for 30 min at which point the reaction mixture had become clear. Reaction mixture 2: In a second round bottom flask CuSO$_4$$\bullet5$H$_2$O (223 mg, 0.89 mmol) is carefully mixed in water (1 mL) with KCN (242 mg, 3.72 mmol) dissolved in water (1mL) keeping the temperature below 20$\degree$C cooling with ice.  To reaction mixture 2 is slowly added reaction mixture 1 while keeping the reaction basic by adding Na$_2$CO$_3$ (sat.). After the addition is complete the reaction is left stirring 1 hour at room temperature. CH$_2$Cl$_2$ (30 mL) and Et$_2$O (50 mL) are added and the organic phase was washed with water (2$\times$20 mL), brine (20 mL) and dried using MgSO$_4$. The solvents were removed under reduced pressure and the crude reaction mixture was purified by column chromatography eluting with pentane:CH$_2$Cl$_2$ 3:1 to afford the title compound as a colorless solid (141.8 mg, 50\%). $^1$H NMR (400 MHz, CDCl$_3$) $\delta$ (ppm) 7.8 (s, 2H), 7.11-7.05 (m, 2H), 6.92 (tt, 1H, J = 8.6, 2.3 Hz). $^{13}$C NMR (400 MHz, CDCl$_3$) $\delta$ (ppm) 163.7 (249.1, 12.7 Hz), 145.2, 140.0 (t, J = 9.4 Hz), 130.5, 127.5, 118.4, 115.9, 110.7 (dd, J = 18.9, 7.7 Hz), 105.3 (t, J = 25.0 Hz). $^{19}$F NMR (376 MHz , CDCl$_3$) $\delta$ (ppm) -107.7 (t, J = 8.0 Hz). GCMS. C$_{13}$H$_5$F$_2$NBr$_2$ Calculated:372.99. Found: 373 (100), 213 (68), 186 (15), 106 (15).

%\bibliography{ref}
%

\end{document}